# Impact of high *Q* on ILC250 upgrade for record luminosities and path toward ILC380

H Padamsee[1,2], A. Grasselino[1], S. Belomestnykh[1], and S. Posen[1]

[1]*Fermi National Accelerator Laboratory, Batavia, Illinois, 60510, USA*
[2]*Cornell University, Ithaca, New York, 14853, USA*

*Abstract*

While the Japanese community deliberates on a decision for the ILC-based Higgs Factory at 250 GeV [1], the European Strategy discussions [2] include focus on the CERN-initiated option of FCC-ee [3]. Arguments for FCC-ee are the high luminosity possible ($1.7\times10^{35}$) with high currents in rings, simultaneous availability of two detectors, familiar technology of colliding rings versus linear colliders, as well as a far-future hadron collider at 100 TeV in the same tunnel. A significant challenge for FCC-ee will be large cost of a 100 km tunnel and associated components. Upon conclusion of the Conceptual Design Report [3] the total cost for FCC-ee (Higgs Factory) is stated to be 10.5 B CHF (including tunnel, but not including manpower or detectors).

In comparison, the stated luminosity of ILC250 is $1.35\times10^{34}$ but at a factor of two lower cost of about 5.5 B ILC Units (ILCU, equivalent $) for option A' [1], also including tunnel but not including detector and manpower [1]. The ILC cost is based on a mature Technical Design Report. The availability of polarized beams increases the effective ILC luminosity by a factor of 2.5 [4], compensating for the single detector, versus two detectors for FCC-ee. Polarization is one of the merits of linear colliders over circular machines and its impact is judged from the effect on Higgs coupling measurements. We note that the ILC already discusses [5] a luminosity upgrade option (with a somewhat different parameter approach than ours) to raise ILC luminosity by a factor of 2, which will make the effective luminosity with polarization $6.8\times10^{34}$ ($1.35 \times 2 \times 2.5$), but still not competitive with FCC-ee, as outlined in their CDR.

In this paper, we address the possibility of upgrading the ILC250 luminosity to $8.1\times10^{34}$, so that with the polarization feature, the effective luminosity will be $2.0\times10^{35}$ to compete with the FCC-ee luminosity and two detectors. The additional cost of the higher luminosity option will be about 2.2 B ILCU. The total cost for the ILC high luminosity machine will therefore be about 7.7 B versus FCC-ee 10.5 B. The AC power (267 MW) to operate the ILC luminosity upgrade will also be less than the AC power for FCC-ee (300 MW). Even with a modest quality factor *Q* of $1\times10^{10}$ for SRF cavities, the total cost of the upgrade will be 2.5 B ILCU additional over ILC250 baseline. We expect that, if approved, ILC250 will first be built at the baseline luminosity, operated for many years at this luminosity, and later upgraded to the high luminosity option. A significant part (RF power and cryo-power) of the additional cost for the luminosity upgrade overlaps with the expected additional costs for anticipated energy upgrade paths.

A second ILC upgrade discussed in this paper will be to the higher energy Top Factory at 380 GeV. We also estimate the additional cost of this upgrade (1.5 B ILCU).

*Upgrade to high luminosity*

On May 8, 2019, Fermilab held a one-day workshop to discuss strategies and costs for higher ILC250 luminosities. The agenda and talks at the meeting can be found at

https://indico.fnal.gov/event/20759/

The preliminary workshop conclusions were summarized and shared with participants at the European Strategy Meeting in Granada on May 13, 2019 [2]. Some of the questions raised have been further addressed in this paper.

Substantial R&D, engineering and other studies have been accomplished to reach attractive physics parameters and provide mature technical designs for ILC 250 and ILC 500. By contrast, this study is in a very early exploration stage.

In our approach to raising ILC250 luminosity we chose to keep the main beam parameters the same as for ILC250 so that the final collision spot size and the detector backgrounds of ILC remain unchanged. We rely on the benefits of higher quality factor $Q$ performance of SRF cavities ($2 \times 10^{10}$ at 31.5 MV/m), which opens the option of increasing the RF pulse length (and so the beam-on duty cycle) allowing the population of the RF pulse with twice the number of bunches (1,312 to 2,624) at the same bunch spacing in the linac as for the ILC baseline which also helps to better preserve emittance in the linac. We increase the repetition rate of the pulses from 5 Hz to 15 Hz to give a total increase of factor of 6 in average beam power (from 5.3 MW to 31.5 MW) and so also the luminosity. The additional costs were determined from the ILC250 baseline costs, see Table 1.

*Refrigerator considerations.*

Due to the longer pulse length (by 45%) and the higher repetition rate (×3) the dynamic heat load increases, but the factor of two higher $Q$ helps to keep the total heat load (including static) increase to near 2 times the ILC baseline. Hence the cost for refrigeration increases by about a factor of two from the baseline. We note that this additional cryogenic power for the luminosity upgrade is also needed for the later energy upgrades to 380 GeV (Top Factory) and to 500 GeV as in the ILC-TDR. For the lower $Q$ of $1 \times 10^{10}$, the cryogenic power would rise from 17 kW become 55 kW. Clearly the higher $Q$ has a large impact.

*RF power considerations.*

The longer RF pulse length and duty factor keeps the total RF peak power requirement the same, but the average RF power per klystron/modulator increases due to higher repetition rate and longer pulse length. We propose to double the number of 10 MW klystrons but use all the klystrons at 5 MW peak power for the luminosity upgrade. The RF will be distributed differently (to fewer cavities per klystron) in order to achieve the same 31.5 MV/m gradient. In this case the average power per klystron will be 175 kW, with average power per klystron window of 88 kW. The ERL injector at Cornell has klystrons operating at 1.3 GHz with the individual window power limit of 135 kW CW [6]. Therefore, operating with the average power of 88 kW per klystron window should be possible. The cost of RF doubles for the luminosity upgrade, see Table 1. ILC will need these additional klystrons at 10 MW for future energy upgrades, as is the case with the additional refrigeration capacity.

*Damping Rings*

The simplest (and most expensive) way to deal with double the number of bunches is to double the number of damping rings (and the costs) during the shut-down to install the high luminosity upgrade. (Note that ILC TDR already includes the option of a second positron ring above the first one.) To deal with 15 Hz rep rate the damping time needs to be reduced by a factor of 1.5: Recall that one of the DR operating modes described in the TDR is for 10 Hz operation. The reduced damping time can be accomplished by increasing the wiggler length, or by installing new wigglers with higher field and shorter periods as under development for CLIC [7]. The ILC TDR design is based on 2.2 T wiggler field with 30 cm period. The CLIC CDR design has 2.5 T wigglers with 5 cm period. Under development at CERN are wigglers based on $Nb_3Sn$ with 3 - 4 T field [8]. CLIC has a repetition rate of 50 Hz, and needs 2 ms damping time, compared to 67 ms damping time for the high luminosity ILC.

There are also additional conventional DR facility costs which are included in the upgrade cost estimate. Layouts still need to be developed. Clearly the improved DRs needs much further study and engineering to advance to a technical design.

*Positron source*

The ILC baseline designs for both the electron and positron sources are specified for the production of 2560 bunches required for the luminosity upgrade (see Chapter 4 and Chapter 5 of ILC TDR). The 3× higher repetition rate will impose substantially higher power demands on targets and dumps. A direct way to address the higher powers would be to install 3 parallel systems. Space would have to allow for such a future upgrade during installation of the 250 GeV baseline. This requires positron source costs to increase by a factor of 3.

An alternative is to explore the normal conducting linac option for positron generation described in the ILC 250 GeV TDR. The disadvantage here would be to lose the polarization of the positrons, which may introduce some systematic errors into Higgs coupling measurements [9]

Table 1 shows the cost breakdown comparison of the major systems for ILC250 and ILC250-lumi-upgrade. The incremental cost for the luminosity upgrade is about 2.2 B ILCU. Even with a modest $Q$ of $1 \times 10^{10}$ the total cost of the upgrade will be 2.5 B ILCU additional over the ILC250 baseline.

*Beam Dump*

Due to 3× higher beam power accompanying the higher repetition rate the cost of the beam dumps would also have to be increased by a factor of 3, as shown in Table 1.

Table 1: Changes to ILC250 machine parameters when increasing luminosity by a factor of 6 relative to the TDR. Two $Q$ values are considered for the cavities: the baseline value of $1\times10^{10}$ and one considering the significant progress in SRF R&D since the 2012 TDR allowing for $2\times10^{10}$. Costs are given in B ILCU which is approximately equivalent to billions of US$.

|  | TDR Baseline | 6× Luminosity upgrade, with $Q = 1\times10^{10}$ | 6× Luminosity upgrade, with $Q = 2\times10^{10}$ |
|---|---|---|---|
| **Energy [GeV]** | 250 | 250 | 250 |
| **Luminosity [$\times10^{34}$]** | 1.35 | 8.1 | 8.1 |
| **Total capital cost (no labor) [B ILCU]** | 5.5 | 8.0 | 7.73 |
| **Total AC power [MW]** | 132 | 286 | 267 |
| **Cyomodules (including cavities) [B ILCU]** | 1.93 | 1.93 | 1.93 |
| **Conventional Facilities, CF (all) [B ILCU]** | 1.43 | 1.63 | 1.63 |
| **Refrigeration system [B ILCU]** | 0.5 | 1.3 | 1.0 |
| **High power RF (linac only) [B ILCU]** | 0.6 | 1.2 | 1.2 |
| **Damping ring [B ILCU]** | 0.33 | 0.66 | 0.66 |
| **Positron source [B ILCU]** | 0.23 | 0.69 | 0.69 |
| **Beam Dumps [B ILCU]** | 0.07 | 0.21 | 0.21 |
| **Other systems (not including CF) [B ILCU]** | 0.41 | 0.41 | 0.41 |
| **Tunnel length [km]** | 20 | 20 | 20 |
| **Gradient [MV/m]** | 31.5 | 31.5 | 31.5 |
| **$Q$** | $1\times10^{10}$ | $1\times10^{10}$ | $2\times10^{10}$ |
| **Repetition rate [Hz]** | 5 | 15 | 15 |
| **Number of bunches** | 1,312 | 2,624 | 2,624 |
| **Beam power [MW]** | 5.3 | 31.5 | 31.5 |
| **Total RF pulse length [ms]** | 1.618 | 2.35 | 2.35 |

*High Q development*

Given the remarkable progress described below in high $Q$ at high gradients, it is reasonable to anticipate that a robust procedure for 9-cell ILC cavities will become available in a few years and in time for ILC 250 baseline installation. Hence it will not be necessary to re-treat the cavities for the high $Q$ needed for the luminosity upgrade. Another concern about high $Q$ is the possible degradation from field emission due to accidental contamination. Progress is forthcoming on this front from two arenas: high pulsed power processing [10] and more recently, plasma processing [11]. One can anticipate a routine cavity conditioning protocol to restore $Q$ values to needed levels.

Fermilab has developed two new approaches to high $Q$ and proved these with many single cell cavities at 1.3 GHz. The first approach is *nitrogen infusion* [12]. Figure 1 shows the typical results. Here the $Q$ of $2\times10^{10}$ is achieved at 31.5 MV/m. Work is in progress at KEK and JLab to reproduce these results [13]. JLab has demonstrated success at 140 C [14]. One challenge is that the infusion method is sensitive to furnace cleanliness and may be difficult to implement on a wide scale.

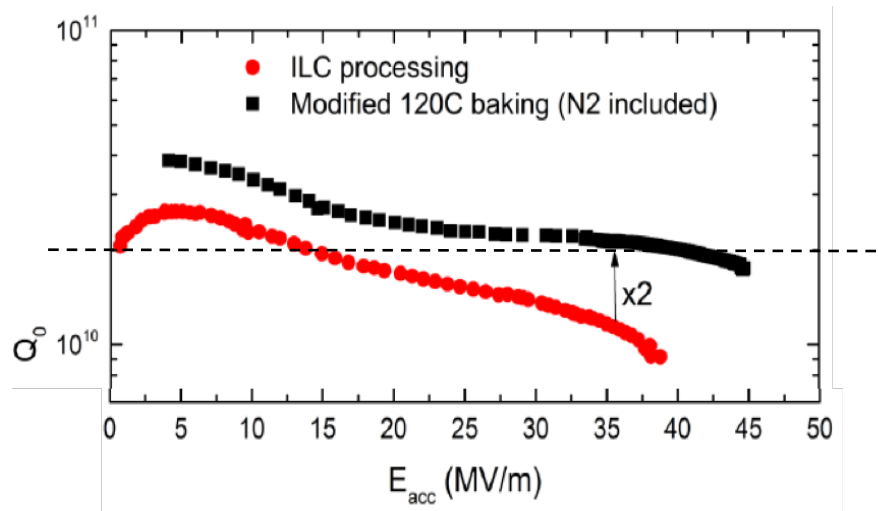

Figure 1: High $Q$ at high gradient achieved by nitrogen infusion process at Fermilab. $Q$ above $2\times10^{10}$ is reached at 31.5 MV/m.

The second approach [15], called *two-step baking*, is simpler than nitrogen infusion and is established with many single cells. During the final preparation of the cavity, the bake of 120ºC for 48 hours is modified to add a 6 hour bake at 75ºC. Figure 2 shows the encouraging results. Basic R&D is in progress to understand the fundamental mechanisms at work for these new treatments [16].

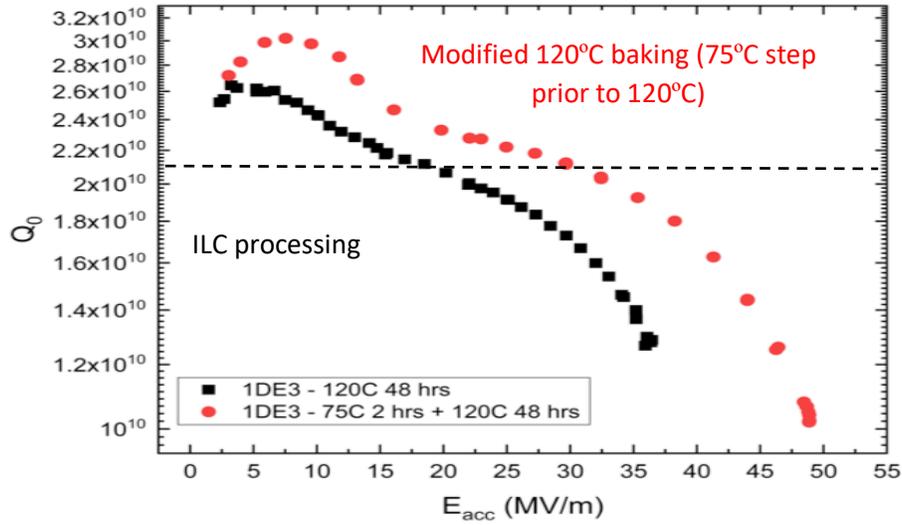

Figure 2: High $Q$ at high gradient achieved by "75/120°C bake" process at Fermilab. $Q$ above $2\times10^{10}$ is reached at 31.5 MV/m.

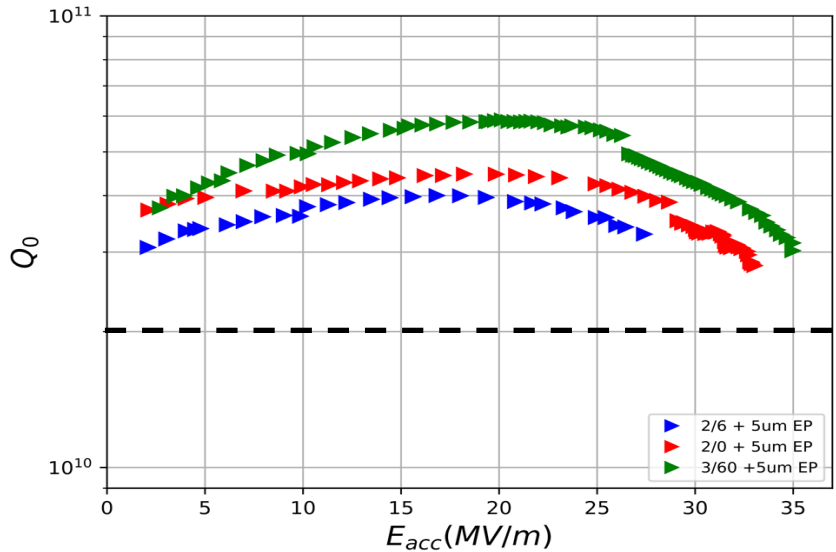

Figure 3: High $Q$ at high gradient achieved by 800°C nitrogen doping at Fermilab and JLab. $Q$'s above $2\times10^{10}$ are reached at 31.5 MV/m. $Q$'s exceed $3\times10^{10}$.

Both Fermilab and JLab are working on high $Q$'s at high gradients for LCLS-II-HE using 800°C nitrogen doping followed by light EP which helps to remove any furnace contaminants. There are two recipes that give encouraging results, one called 2/0 (FNAL) and the other 3/60 (JLab). In the first case there is 2 minutes of 800°C N doping with no annealing, and the second procedure is 3 minutes of N-doping followed by 60 minutes annealing. Both procedures give $Q$ values above $2\times10^{10}$ at 31.5 MV/m. The 2/0 recipe has also been applied successfully to a 9-cell cavity, with results shown in Figure 4.

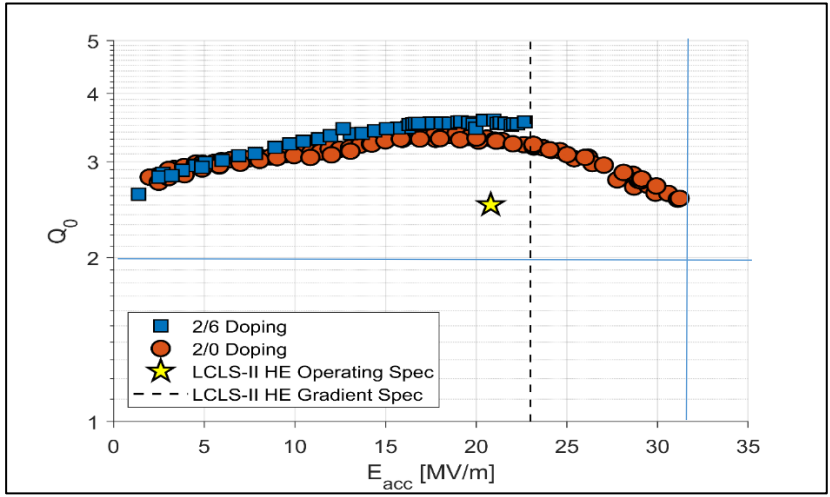

Figure 4: High $Q$ at high gradient achieved by 800ºC nitrogen doping, 2/0 recipe, at Fermilab with a 9-cell cavity. $Q$ above $2\times10^{10}$ is reached at 31.5 MV/m.

*ILC 380 GeV Top Factory upgrade*

The ILC TDR includes upgrade paths to 500 GeV and 1 TeV. At the May 2019 workshop, attendees also examined the parameters, costs and AC power for the ILC 380 GeV Top Factory to compare with the FCC-ee Top Factory option. The availability of additional damping rings and positron sources from the high luminosity upgrade can be exploited to raise the Top Factory luminosity to $4\times10^{34}$ with twice the number of bunches per RF pulse, repetition rate of 5 Hz, with AC power of 188 MW. This should be compared to FCC-ee (Top Factory) luminosity of $1.6\times10^{34}$ with AC power of 365 MW. The big advantage of the ILC path to the Top Factory is the substantially lower AC power.

While the baseline ILC250 and its high luminosity upgrade operate for 10-15 years, we expect the SRF gradient and $Q$ technology to continue to advance. Based on single cell results available today, we can expect that the additional 3,000 cavities needed for 380 GeV will be able to reach 40 MV/m [12,15] at $Q$'s of $2\times10^{10}$. R&D needs to be carried out to bring these results from single cell to 9-cell cavities. The average gradient over the entire 380 GeV accelerator would become 34.4 MV/m at an average $Q$ of $1.34\times10^{10}$.

If the high luminosity upgrade is already in place, the additional RF and additional refrigeration needed for 380 GeV are already in place, with spare capacity for future energy upgrades. The RF power distribution of 10 MW klystrons installed for the high luminosity upgrade would be re-adjusted to achieve the 40 MV/m cavity gradient and average power per klystron. Therefore, the additional cost for 380 GeV upgrade would be for the new cryomodules (0.72B) and the additional conventional construction (0.77B) totaling 1.5B on top of the ILC250-lumi-upgrade stage.

For both the high luminosity upgrade (+2.2B) and the 380 GeV upgrade (+1.5B), the total additional cost is 3.7B over the baseline of 5.5B. The grand total ILC baseline + high luminosity + Top Factory is 9.2 B as compared to FCC-ee 11.6B for Higgs factory + Top Factory [3].

*Alternative options (beam parameter changes)*

Given the technical challenges described for damping rings and positron sources due to the 6× higher beam power (2× number of bunches and 3× repetition rate), our approach to keep the beam parameters unchanged is possibly too conservative. The final focus collision spot size of 7.7 nm is still significantly larger than normal conducting avenues to future linear colliders. In addition, the calculated backgrounds are also respectably low, so there are other avenues to explore by reducing the collision spot size. There are many options. One example (to judge the impact on backgrounds) is to lower final spot size from 7.6 nm to 6 nm by reducing $\beta x/\beta y$ from 13 mm/0.4 mm to 10 mm/0.26 mm, but with increased bunch length from 0.3 to 0.4 mm. The beamstrahlung parameter increases from 0.038 to 0.046, and the number of incoherent pairs increases from $1.3 \times 10^3$ (for ILC 250 TDR baseline) to $4.2 \times 10^3$. The luminosity with 5 Hz repetition rate, and 2,624 bunches remains competitive with FCCee. We will continue to explore similar avenue in the future.

*Final Remarks*

A primary advantage of the ILC path for lepton colliders is that the ILC design is at present in the mature TDR stage whereas the FCC-ee needs several years of development from its current CDR stage to advance to TDR with bottoms-up cost estimates. Avenues for 10% cost reduction for ILC are under exploration at many laboratories [17]. The technology of ILC is highly mature and significantly industrialized with many laboratories around the world having acquired the expertise. A 10% ILC prototype (23 GeV – 800 cavities) exists with the EXFEL facility at DESY [18]. High $Q$ ($2.7 \times 10^{10}$) technology has been adopted for the LCLS-II at 4 GeV (250 cavities) [19] and for LCLS-II-HE [20] at 8 GeV (+200 cavities), both accelerators planning CW operation. A variety of accelerators (ESS, PIP-II, SHINE) are under construction to install more than 1,000 SRF cavities over the next 5 years [21]. The SRF technology has a strong established base worldwide at laboratories and industry.

*Acknowledgements*

Thanks to all the participants at the May 2019 workshop, as well as Pushpa Bhat, Nigel Lockyer and Nick Walker for discussions.